\documentclass{article}
\usepackage{spconf,amsmath,graphicx, amsmath}
\usepackage{hyperref}

\usepackage{enumitem}
\setlist{nosep, leftmargin=14pt}

\usepackage{mwe} 
\usepackage{multirow}

\title{MrRegNet: Multi-resolution Mask Guided Convolutional Neural Network for Medical Image Registration with Large Deformations}
\name{Ruizhe Li$^{1}$ \qquad Grazziela Figueredo$^{2}$ \qquad Dorothee Auer$^{2}$ \qquad Christian Wagner$^{1}$ \qquad Xin Chen$^{1}$}
\address{$^{1}$IMA/LUCID Group, School of Computer Science, University of Nottingham, UK\\
        $^{2}$School of Medicine, University of Nottingham, UK}
\begin{document}
%
\maketitle
\begin{abstract}

Deformable image registration (alignment) is highly sought after in numerous clinical applications, such as computer aided diagnosis and disease progression analysis. Deep Convolutional Neural Network (DCNN)-based image registration methods have demonstrated advantages in terms of registration accuracy and computational speed. However, while most methods excel at global alignment, they often perform worse in aligning local regions. To address this challenge, this paper proposes a mask-guided encoder-decoder DCNN-based image registration method, named as MrRegNet. This approach employs a multi-resolution encoder for feature extraction and subsequently estimates multi-resolution displacement fields in the decoder to handle the substantial deformation of images. Furthermore, segmentation masks are employed to direct the model's attention toward aligning local regions. The results show that the proposed method outperforms traditional methods like Demons and a well-known deep learning method, VoxelMorph, on a public 3D brain MRI dataset (OASIS) and a local 2D brain MRI dataset with large deformations. Importantly, the image alignment accuracies are significantly improved at local regions guided by segmentation masks. Github link: \href{https://github.com/ruizhe-l/MrRegNet}{https://github.com/ruizhe-l/MrRegNet}.

\end{abstract}
\vspace{-0.1cm}
\section{Introduction}
\label{sec:intro}

Deformable image registration is a technique that transforms a source image into the space of a target image, facilitating their comparison and analysis. This method is widely used in various fields, especially medical image alignment for disease diagnosis and prognosis. Despite decades of method development in image registration, some issues remain to be addressed. One of the key challenges is to cope with large image deformations. 

Traditional methods use iterative optimization frameworks, requiring parameter tuning and significant time for 3D medical image registration. In contrast, deep learning methods, such as deep convolutional neural networks (DCNN), have shown impressive performance in medical image registration \cite{balakrishnan2018unsupervised,de2019deep}. They require training with example images, offering faster registration for unseen images, typically taking seconds, and ensuring high accuracy. However, they often struggle with aligning images featuring significant deformations.

Traditional image registration methods address large deformations through a multi-resolution approach, which inspired the development of multi-resolution deep learning techniques. Hering et al. \cite{hering2019mlvirnet} introduced a multi-resolution image registration framework using U-Net-based networks \cite{ronneberger2015u} to estimate displacement fields at various resolutions. This method involved training networks incrementally from coarse to fine scales, utilizing down-sampled target images and warped source images from lower resolution networks as inputs. In contrast, Mok et al. \cite{mok2020large} presented a Laplacian Pyramid method employing lightweight encoder-decoder networks with skip connections to handle large deformations. These networks focused on different image resolutions, and higher resolution networks used the source image, target image, and up-sampled displacement fields from lower resolution networks as inputs. Similarly, several U-Net based multi-resolution methods were proposed that combined feature maps from different resolutions \cite{kim2021cyclemorph} \cite{li2022mdreg}. These approaches significantly outperformed single-resolution and traditional methods for large deformation image registration.

Building on the multi-resolution concept, this paper introduces a multi-resolution DCNN method for 2D and 3D medical image registration. Unlike existing methods, this approach uses a single encoder to extract features at different scales, with the decoder generating residual displacement fields for different resolutions. The finest scale displacement field is estimated by sequentially combining up-sampled displacement fields from coarser scales. The main contributions of this work are summarized as follows. (1) A single encoder provides a lightweight model and the flexibility for adjusting the number of resolutions. (2) It no longer requires a warped source image as input to higher resolution layers, which requires shorter training time. (3) Efficient and effective diffeomorphic deformation is achieved through a residual displacement field estimation. (4) It incorporated a mask-guided loss term to improve local image alignment.

\begin{figure}
    \centering
    \includegraphics[width=\columnwidth]{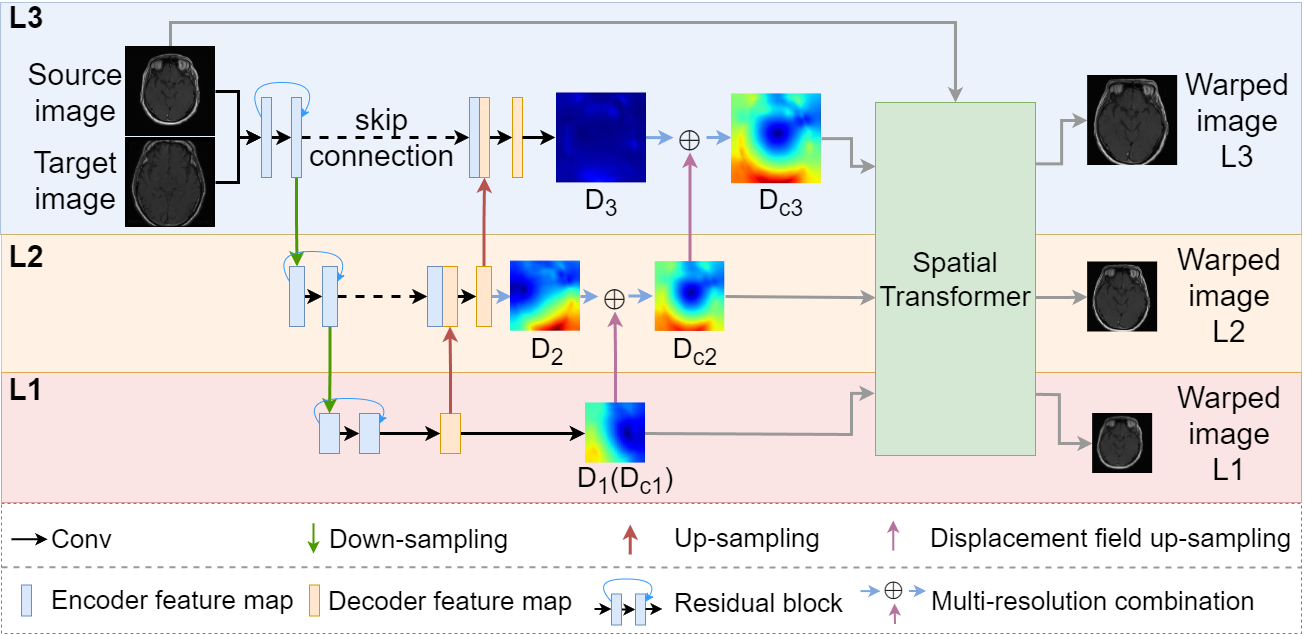}
    \vspace{-0.5cm}
    \caption{Overview of the proposed multi-resolution image registration framework with 3 levels. }
    \label{fig:framework}
    \vspace{-0.4cm}
\end{figure}

\vspace{-0.2cm}
\section{Method}
\label{sec:methods}

\subsection{Model Architecture}
\label{ssec:methods-model}

An overview of the proposed image registration framework is shown in Fig. \ref{fig:framework}. The network consists of an encoder and a decoder, and each has $K$ levels ($L_1$,…,$L_K$) of scales. Fig. \ref{fig:framework} is an example framework with 3 levels. In both the training and inference processes, pairs of source and target images are merged into two-channel images, which are subsequently fed into the network. The key components are introduced below. 

\textbf{Encoder-decoder:} in the encoder, each layer has a residual block consisting of two 3×3 convolutional operators (3×3×3 for 3D) with a stride of 1, followed by Leaky ReLU as the activation function \cite{maas2013rectifier}. A 1×1 (1×1×1 for 3D) convolutional operator with a stride of 2 is applied to down-sample the feature maps between two consecutive levels. In the decoder, each level consists of one 3×3 (3×3×3 for 3D) convolutional operator with a stride of 1 and one 3×3 (3×3×3 for 3D) de-convolutional operator with a stride of 2 as up-sampling layer. Leaky ReLU is used after both of them as the activation function. A skip connection is applied to concatenate the feature maps of the encoder to the decoder at the same level.

\textbf{Residual Displacement Field:} in Fig. \ref{fig:framework}, a 3x3 (3x3x3 for 3D) convolutional operator was employed at each decoder level to estimate displacement fields ($D_i$) corresponding to the image resolution. During training, we first estimate the displacement field $D_{c1}$ for the lowest resolution in $L_1$. In $L2$, we calculate $D_2$ by decoding and add it to the up-sampled $D_{c1}$ to form the final $D_{c2}$ in $L2$. To up-sample $D_{c1}$, we double its size with linear interpolation and scale the displacement values by 2. This process is repeated at higher levels to combine displacement fields from the previous level. Each $D_i$ acts as a residual displacement map, facilitating more efficient learning and enhancing diffeomorphic deformation.

\textbf{Spatial Transformer and Image Warping:} to enable unsupervised image registration, the spatial transformer layer \cite{jaderberg2015spatial} is used in this framework. We input randomly paired source and target images into the network, where the encoder and decoder jointly create a displacement field. Subsequently, the spatial transformer layer is employed to transform the source image based on the displacement field, resulting in a warped image. The model iteratively fine-tunes its weights to maximize the similarity between the warped source image and the target image.

\subsection{Model Training}
Given a source image $S$ and a target image $T$, the objective of image registration is to transform $S$ so that it can be aligned with $T$. The transformation in this case is represented by a displacement field $D$, which is used to warp the source image (denoted as $f_D(S)$). The image registration network is then optimized based on a similarity measurement between the target image and the warped source image. A widely used metric, global normalized cross-correlation (GNCC) \cite{avants2008symmetric}, is employed as the similarity loss. Additionally, the loss function incorporates a smoothness regularization term \cite{balakrishnan2019voxelmorph} to regularize the displacement field. It is important to highlight that both of these components are computed at every resolution level. Therefore, the loss function for the registration model is as follows:
\vspace{-0.1cm}
\begin{equation}
    \label{eq:1}
    L_{reg} = \frac{1}{K}\sum_{i=1}^{K}(GNCC(f_{D_{ci}}(S), T) +\lambda \left \| \bigtriangledown D_i \right \|)
\end{equation} 
\vspace{-0.1cm}
where $\bigtriangledown D_i$ is the smoothness term that denotes the approximate spatial gradients of displacement $D_i$ using differences between neighboring pixels (voxels).



In the context of medical image analysis, it is very common to focus on a small region of interest (ROI) (e.g. brain stem). In the task of image registration, a globally aligned image pair does not guarantee a local optimal alignment. Hence, in this work, a mask guided term is introduced to the multi-resolution framework during model training. In each training iteration, the mask of the source image $S_{mask}$ is transformed using the displacement fields at different resolutions to produce warped masks ($f_{D_{ci}}(S_{mask})$). The target mask is then down-sampled to match the corresponding image sizes, and the final loss is computed as the average of the similarities of all $K$ scales. To address class imbalance, soft dice loss is used here instead of cross-entropy loss:
\vspace{-0.1cm}
\begin{equation}
    \label{eq:5}
    DSC(x,y) = \frac{1}{K}\sum_{i=1}^{K} \frac{2 \|x \cdot y\|}{\|x\|^2 + \|y\|^2}
\end{equation}
\vspace{-0.1cm}
where $x$, $y$ represent the warped mask $f_D(S_{mask})$ and the target mask $T_{mask}$.

Finally, the loss function for model training incorporating the mask-driven loss is represented by Eq. (\ref{eq:6}). The network is then trained based on a training set that contains random pairs of source and target images, as well as their corresponding ROI masks as an optional setting. 
\vspace{-0.3cm}
\begin{multline}
    \label{eq:6}
    \underset{D}{arg \; min} \, \frac{1}{K}\sum_{i=1}^{K}(GNCC(f_{D_{ci}}(S), T) + \\
    DSC(f_{D_{ci}}(S_{mask}), T_{mask})+\lambda \left \| \bigtriangledown D_i \right \|)
\end{multline}

Note that the ROI masks are only used in the training process to guide the model learning focusing on the masked region, while masks are not needed during the model inference for aligning unseen images. 

\vspace{-0.2cm}
\section{Method Evaluation}
\label{sec:experiments}

\subsection{Materials and Experiments}
\label{ssec:dataset}

The proposed method was evaluated on two brain MRI datasets:

\textbf{Public 3D Brain MRI Dataset:} This dataset comprises 414 T1-weighted brain scans from the OASIS dataset \cite{marcus2007open}, resized to $96\times96\times96$ for consistency. It includes four manually annotated anatomical regions: cortex, subcortical gray matter, white matter, and cerebrospinal fluid.

\textbf{Local 2D Brain MRI Dataset:} This dataset consists of 820 T1-weighted slices with manual annotations of mid-brain regions. These slices originate from various brain regions of different subjects, acquired using different MRI scanners. Images were resized to $256\times256$ for consistency. This dataset is more challenging due to significant spatial variations across subjects.

For the 3D dataset, we split it into training (60\%), validation (5\%), and test sets (35\%) - roughly 244, 20, and 150 images. The 2D brain dataset was divided into training (75\%), validation (5\%), and test sets (20\%) - approximately 600, 40, and 180 images. 

During training, each image served as both the target and source image in a randomized manner. To evaluate the methods, we used 5 randomly selected source images from the test set and the remaining images as the target, resulting in 875 paired 2D images and 725 paired 3D images for testing. Hyper-parameters for MrRegNet were tuned using the validation dataset with different settings for the 2D and 3D datasets.

We compared our proposed method, MrRegNet, with a widely used traditional method (Demons \cite{vercauteren2009diffeomorphic}) and a well-known deep learning method VoxelMorph \cite{balakrishnan2019voxelmorph} using their official implementations. In addition to assessing MrRegNet's performance, we conducted extra experiments to evaluate the advantages of the mask-guided loss term, named MrRegNet-M. The 2D brain data has annotations for one class in the central brain region, while the 3D brain data includes annotations for four classes distributed throughout the brain. These diverse datasets test the proposed method in various scenarios (i.e. binary/multi-class, small/large masks).

\subsection{Parameter Settings}

On the 2D dataset, for MrRegNet, the learning rate was set to 0.001, and the training epochs was 200. The batchsize was set to 10. The number of levels was determined based on the image size, resulting in 5 levels. The smoothness term weight ($\lambda_i$) varies dynamically across levels, ranging from 128 to 8 and halved at each level ($\lambda_1 = 128$, $\lambda_2 = 64$, $\lambda_3 = 32$, $\lambda_4 = 16$, $\lambda_5 = 8$). This approach assigned higher weights to lower resolution levels to enforce higher levels of smoothness in coarse levels. 

On the 3D dataset, the learning rate is reduced to 0.0001 to facilitate smoother training. The training epochs was 200. The batchsize was set to 1. A 4-level network structure was employed according to the image size. The weights assigned to the smoothness term were also dynamic, ranging from 16 to 2, with $\lambda_1 = 16$, $\lambda_2 = 8$, $\lambda_3 = 4$, and $\lambda_4 = 2$. 

\vspace{-0.1cm}
\subsection{Results}
\label{ssec:experiments-results}

The evaluation utilized global normalized cross-correlation (GNCC) and structural similarity index (SSIM) for measuring similarity between warped and target images. Local alignment quality was assessed using Dice coefficient (DSC) and Hausdorff Distance (HD) on annotated regions between warped and target images \cite{mok2020large} \cite{li2022mdreg}. The rate of non-positive Jacobian determinant values ($\|J_D\|\le 0$) evaluated diffeomorphic properties, with lower rates indicating better performance. As baseline, the GNCC, SSIM, DSC and HD between the source image and the target image before registration were reported.

\vspace{-0.3cm}
\subsubsection{2D Brain MRI Data}
\vspace{-0.6cm}
\label{sssec:result-2d}

\begin{table}[!htb]
    \centering 
    \caption{Quantitative evaluation on the 2D dataset. Mean $\pm$ standard deviation is reported for each evaluation metric.}
    \label{table:2d}
    \resizebox{\columnwidth}{!}{
    \begin{tabular}{|l|l|l|l|l|l|}
        \hline
        \textbf{Method} & \textbf{GNCC} & \textbf{SSIM} & \textbf{DSC} & \textbf{HD} & $\boldsymbol{\|J_D\|\le 0}$ \\ 
        \hline
        Baseline          & 0.39$\pm$0.05 & 0.40$\pm$0.04 & 0.66$\pm$0.15 & 17.07$\pm$18.59 & n/a     \\ \hline
        Demons            & 0.75$\pm$0.09 & 0.57$\pm$0.10 & 0.68$\pm$0.17 & 16.21$\pm$19.17 & 0.00$\pm$0.00   \\ \hline
        VoxelMorph      & 0.60$\pm$0.09 & 0.51$\pm$0.06 & 0.72$\pm$0.16 & 16.01$\pm$18.68 & 0.51$\pm$0.10   \\ \hline
        MrRegNet        & 0.80$\pm$0.03 & 0.57$\pm$0.04 & 0.77$\pm$0.10 & 13.83$\pm$18.88 & 0.01$\pm$0.03   \\ \hline
        \hline
        VoxelMorph-M & 0.50$\pm$0.06 & 0.46$\pm$0.04 & 0.87$\pm$0.07 & 16.42$\pm$18.59 & 0.23$\pm$0.08   \\ \hline
        MrRegNet-M & 0.76$\pm$0.04 & 0.56$\pm$0.04 & 0.86$\pm$0.06 & 11.47$\pm$19.24 & 0.07$\pm$0.07   \\ \hline
    \end{tabular}
    }
    
\end{table}

\vspace{-0.6cm}

\begin{figure}[!htb]
    \centering
    \includegraphics[width=\columnwidth]{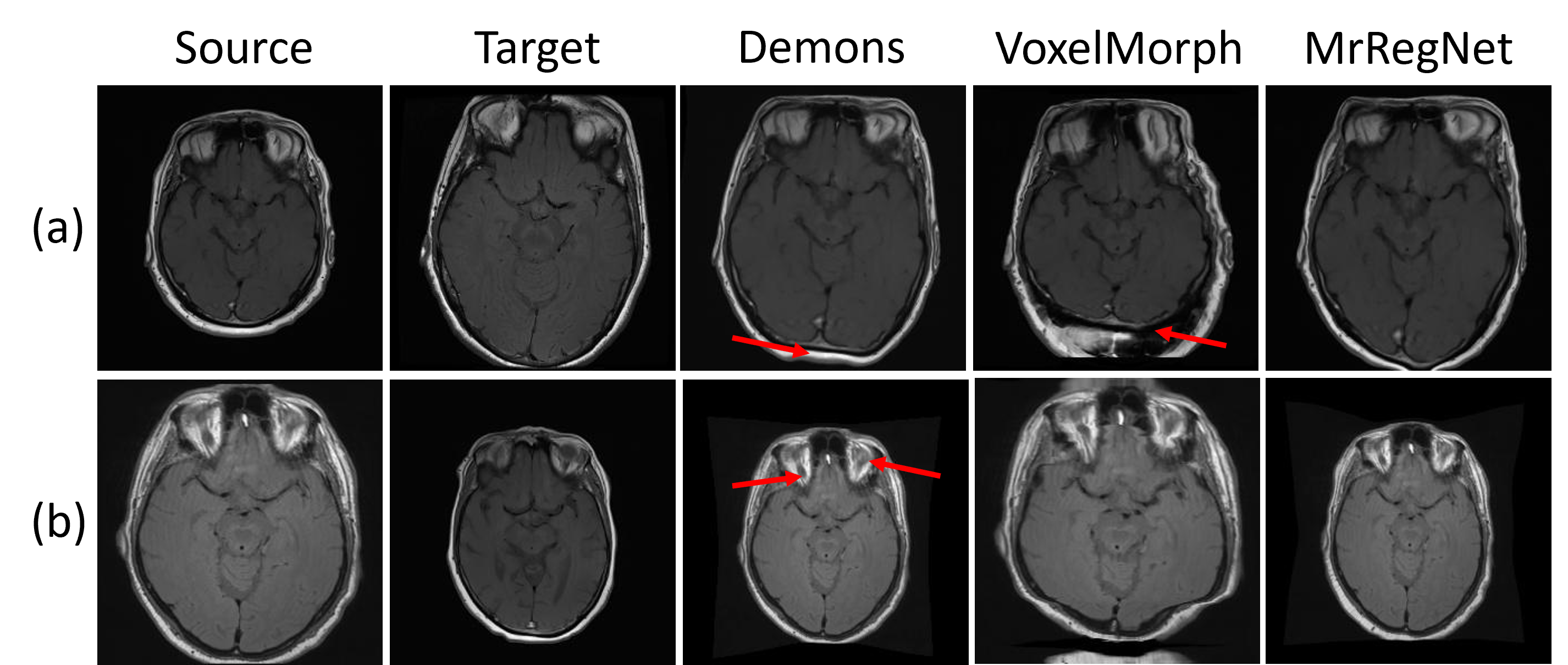}
    \caption{Visualisation results of different registration methods without mask guided loss term on the 2D brain dataset.}
    \label{fig:compare-nomask}
\end{figure}
\vspace{-0.2cm}

Table \ref{table:2d} presents the 2D brain dataset analysis. MrRegNet outperforms others in GNCC and SSIM without a mask-guided component, enhancing local alignment as indicated by DSC and HD improvements. However, adding this component reduces VoxelMorph-M's global performance (GNCC by 0.1, SSIM by 0.05). MrRegNet-M, on the other hand, balances global and local metrics well, with a minor GNCC reduction (0.04) but maintains similar SSIM. It also shows an increase of DSC by 0.09 and achieves better HD.

Furthermore, when comparing VoxelMorph to VoxelMorph-M in Table \ref{table:2d}, the mask-guided component improves DSC but not HD, suggesting it aligns regions but lacks in boundary precision. In contrast, MrRegNet-M improvements both DSC and HD, highlighting the effectiveness of the mask-guided component in combining with MrRegNet. 

Regarding diffeomorphic properties, MrRegNet excels in maintaining a good displacement field. When adding the mask-guided term, VoxelMorph-M performs worse than MrRegNet-M in diffeomorphic measure, and also with a reduced global alignment (i.e. lower GNCC). For MrRegNet, the mask-guided term results in higher non-positive Jacobian determinant values, indicating slightly more folding pixels due to increased local alignment emphasis.

Following quantitative analyses, we performed qualitative assessments by visualizing results in Fig. \ref{fig:compare-nomask}. This figure illustrates three registration examples using Demons, VoxelMorph, and MrRegNet without the mask-guided term. Comparing rows (a) and (b) in Fig. \ref{fig:compare-nomask}, we observe that Demons can handle significant deformations but struggles with accurate boundary alignment (row (a)) and detailed regions, as indicated by the red arrow in the top brain area in row (b). VoxelMorph can align images with substantial deformations but introduces pixel folding issues (row (a), bottom region marked by the red arrow) and faces challenges in correct alignment. MrRegNet produces high quality image alignment even for large deformations.

\vspace{-0.2cm}
\begin{figure}[!htb]
    \centering
    \includegraphics[width=8cm]{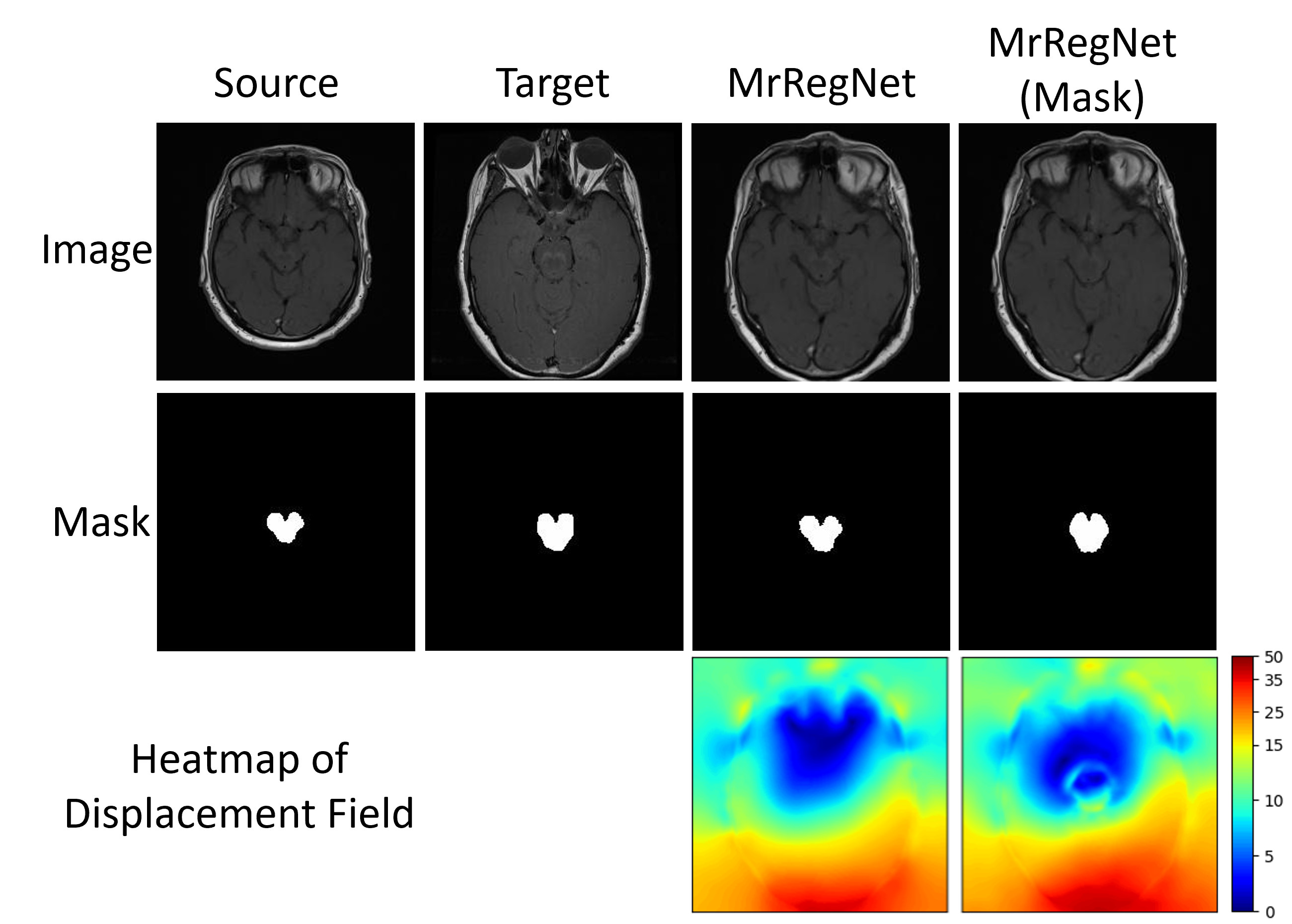}
    \caption{The visualization of MrRegNet with and without mask guided loss.}
    \label{fig:compare-mask-mrregnet}
\end{figure}
\vspace{-0.2cm}

In Fig. \ref{fig:compare-mask-mrregnet}, we visualize an example pair of test images using MrRegNet and MrRegNet-M. We can observe the impact of the mask-guided term by examining the masks and heatmaps of the displacement field in both methods. Compared to MrRegNet, the heatmap of MrRegNet-M shows finer movements in the mid-brain area, emphasizing the mask-guided term's significant effect on local region alignment. The inclusion of the mask-guided term helps capture fine displacements within the masked mid-brain region, resulting in a generated mask that closely resemble the target mask.

\vspace{-0.3cm}
\subsubsection{3D Brain MRI Data}
\vspace{-0.4cm}
\label{sssec:result-3d}

\vspace{-0.2cm}
\begin{table}[!htb]
    \centering
    \caption{Quantitative evaluation on 3D dataset. Mean $\pm$ standard deviation is reported for each evaluation metric.} 
    \resizebox{\columnwidth}{!}{
    \begin{tabular}{|l|l|l|l|l|l|}
     \hline
        \textbf{Method} & \textbf{GNCC} & \textbf{SSIM} & \textbf{DSC} & \textbf{HD} & $\boldsymbol{\|J_D\|\le 0}$ \\ 
        \hline
        Baseline          & 0.67$\pm$0.14 & 0.84$\pm$0.03 & 0.25$\pm$0.11 & 12.89$\pm$4.73 & n/a     \\ \hline
        Demons            & 0.95$\pm$0.02 & 0.96$\pm$0.01 & 0.62$\pm$0.06 & 12.82$\pm$6.43 & 0.01$\pm$0.01   \\ \hline
        VoxelMorph      & 0.93$\pm$0.02 & 0.91$\pm$0.04 & 0.55$\pm$0.06 & 6.59$\pm$2.77 & 0.16$\pm$0.19   \\ \hline
        MrRegNet        & 0.94$\pm$0.02 & 0.92$\pm$0.03 & 0.62$\pm$0.06 & 5.95$\pm$2.56 & 0.13$\pm$0.09   \\ \hline
        \hline
        VoxelMorph-M & 0.92$\pm$0.02 & 0.91$\pm$0.04 & 0.60$\pm$0.11 & 9.17$\pm$3.38 & 0.16$\pm$0.20   \\ \hline
        MrRegNet-M & 0.94$\pm$0.02 & 0.93$\pm$0.03 & 0.67$\pm$0.05 & 5.51$\pm$2.35 & 0.43$\pm$0.15   \\ \hline
    \end{tabular}
    }
    \label{table:3d}
\end{table}
\vspace{-0.3cm}

For the 3D brain dataset, summarized in Table \ref{table:3d}, Demons leads in performance without mask guidance due to uniform intensity across this dataset. Deep learning models like VoxelMorph and MrRegNet, which emphasize anatomical shapes, showed slightly reduced effectiveness in this unique scenario.

Among the deep learning methods, MrRegNet outperformed VoxelMorph in all metrics. Incorporating the mask-guided term into MrRegNet (MrRegNet-M) resulted in significantly improved mean DSC (0.67) and HD (5.51), surpassing all other methods (DSC$\leq$0.62 and HD$\geq$5.95). This improvement did not lead to lower values in the GNCC and SSIM measures, indicating the model's capability to handle both global and local deformations when global deformation is not excessively large.

\section{Discussion and Conclusions}
\label{sec:conclusions}

This study presents a multi-resolution framework for image registration, leveraging deep convolutional neural networks (DCNN) to manage large image deformations. It optimizes residual displacement fields across multiple resolutions, ensuring diffeomorphic deformations via a smoothness term. Our evaluation shows its effectiveness over traditional methods like Demons and advanced learning approaches like VoxelMorph, particularly for 2D brain MRI datasets with substantial deformations, and it performs robustly on 3D brain MRI datasets with multi-class masks. 

The framework's performance is notably enhanced in masked areas by incorporating a mask-guided loss, potentially increasing segmentation accuracy. This advancement allows the framework to serve as a segmentation tool, enabling precise mask warping from source to target image spaces.

In future work, we plan to compare our approach with multiscale models cited in \cite{mok2020large, zhao2019recursive, eppenhof2019progressively, sang2021scale}, focusing on the residual displacement field's impact on diffeomorphic registration, which currently relies on model-based assumptions and empirical evidence, necessitating further theoretical analysis. Additionally, we intend to refine our model with targeted labeled regions to assess if mask guidance enhances alignment across various brain areas.

\vfill
\pagebreak

\section{Compliance with Ethical Standards}
This research study was conducted retrospectively using human subject data made available in the public dataset (OASIS). Ethical approval was not required as confirmed by the license attached with the open access data. The ethical approval of the local brain MRI dataset was in place. The subjects used in this study had consented to be included in this research. All data was anonymized, and the participants’ information cannot be identified from the imaging data.

\section{Conflicts of Interest} No funding was received for conducting this study. The authors have no relevant financial or non-financial interests to disclose.

\bibliographystyle{IEEEbib}
\bibliography{refs}

\begin{thebibliography}{10}

\bibitem{balakrishnan2018unsupervised}
Guha Balakrishnan, Amy Zhao, Mert~R Sabuncu, John Guttag, and Adrian~V Dalca,
\newblock ``An unsupervised learning model for deformable medical image registration,''
\newblock in {\em Proceedings of the IEEE conference on computer vision and pattern recognition}, 2018, pp. 9252--9260.

\bibitem{de2019deep}
Bob~D De~Vos, Floris~F Berendsen, Max~A Viergever, Hessam Sokooti, Marius Staring, and Ivana I{\v{s}}gum,
\newblock ``A deep learning framework for unsupervised affine and deformable image registration,''
\newblock {\em Medical image analysis}, vol. 52, pp. 128--143, 2019.

\bibitem{hering2019mlvirnet}
Alessa Hering, Bram~van Ginneken, and Stefan Heldmann,
\newblock ``mlvirnet: Multilevel variational image registration network,''
\newblock in {\em International Conference on Medical Image Computing and Computer-Assisted Intervention}. Springer, 2019, pp. 257--265.

\bibitem{ronneberger2015u}
Olaf Ronneberger, Philipp Fischer, and Thomas Brox,
\newblock ``U-net: Convolutional networks for biomedical image segmentation,''
\newblock in {\em International Conference on Medical image computing and computer-assisted intervention}. Springer, 2015, pp. 234--241.

\bibitem{mok2020large}
Tony~CW Mok and Albert Chung,
\newblock ``Large deformation diffeomorphic image registration with laplacian pyramid networks,''
\newblock in {\em International Conference on Medical Image Computing and Computer-Assisted Intervention}. Springer, 2020, pp. 211--221.

\bibitem{kim2021cyclemorph}
Boah Kim, Dong~Hwan Kim, Seong~Ho Park, Jieun Kim, June-Goo Lee, and Jong~Chul Ye,
\newblock ``Cyclemorph: cycle consistent unsupervised deformable image registration,''
\newblock {\em Medical Image Analysis}, vol. 71, pp. 102036, 2021.

\bibitem{li2022mdreg}
Hongming Li, Yong Fan, and Alzheimer's Disease~Neuroimaging Initiative,
\newblock ``Mdreg-net: Multi-resolution diffeomorphic image registration using fully convolutional networks with deep self-supervision,''
\newblock {\em Human Brain Mapping}, vol. 43, no. 7, pp. 2218--2231, 2022.

\bibitem{maas2013rectifier}
Andrew~L Maas, Awni~Y Hannun, Andrew~Y Ng, et~al.,
\newblock ``Rectifier nonlinearities improve neural network acoustic models,''
\newblock in {\em Proc. icml}. Atlanta, GA, 2013, vol.~30, p.~3.

\bibitem{jaderberg2015spatial}
Max Jaderberg, Karen Simonyan, Andrew Zisserman, et~al.,
\newblock ``Spatial transformer networks,''
\newblock {\em Advances in neural information processing systems}, vol. 28, 2015.

\bibitem{avants2008symmetric}
Brian~B Avants, Charles~L Epstein, Murray Grossman, and James~C Gee,
\newblock ``Symmetric diffeomorphic image registration with cross-correlation: evaluating automated labeling of elderly and neurodegenerative brain,''
\newblock {\em Medical image analysis}, vol. 12, no. 1, pp. 26--41, 2008.

\bibitem{balakrishnan2019voxelmorph}
Guha Balakrishnan, Amy Zhao, Mert~R Sabuncu, John Guttag, and Adrian~V Dalca,
\newblock ``Voxelmorph: a learning framework for deformable medical image registration,''
\newblock {\em IEEE transactions on medical imaging}, vol. 38, no. 8, pp. 1788--1800, 2019.

\bibitem{marcus2007open}
Daniel~S Marcus, Tracy~H Wang, Jamie Parker, John~G Csernansky, John~C Morris, and Randy~L Buckner,
\newblock ``Open access series of imaging studies (oasis): cross-sectional mri data in young, middle aged, nondemented, and demented older adults,''
\newblock {\em Journal of cognitive neuroscience}, vol. 19, no. 9, pp. 1498--1507, 2007.

\bibitem{vercauteren2009diffeomorphic}
Tom Vercauteren, Xavier Pennec, Aymeric Perchant, and Nicholas Ayache,
\newblock ``Diffeomorphic demons: Efficient non-parametric image registration,''
\newblock {\em NeuroImage}, vol. 45, no. 1, pp. S61--S72, 2009.

\bibitem{zhao2019recursive}
Shengyu Zhao, Yue Dong, Eric~I Chang, Yan Xu, et~al.,
\newblock ``Recursive cascaded networks for unsupervised medical image registration,''
\newblock in {\em Proceedings of the IEEE/CVF international conference on computer vision}, 2019, pp. 10600--10610.

\bibitem{eppenhof2019progressively}
Koen~AJ Eppenhof, Maxime~W Lafarge, Mitko Veta, and Josien~PW Pluim,
\newblock ``Progressively trained convolutional neural networks for deformable image registration,''
\newblock {\em IEEE transactions on medical imaging}, vol. 39, no. 5, pp. 1594--1604, 2019.

\bibitem{sang2021scale}
Yudi Sang and Dan Ruan,
\newblock ``Scale-adaptive deep network for deformable image registration,''
\newblock {\em Medical Physics}, vol. 48, no. 7, pp. 3815--3826, 2021.

\end{thebibliography}

\end{document}